\documentclass[aps,prl,twocolumn,showpacs,superscriptaddress,groupedaddress]{revtex4}  % for review and submission
\usepackage{graphicx}  % needed for figures
\usepackage{dcolumn}   % needed for some tables
\usepackage{bm}        % for math
\usepackage{amssymb}   % for math

\usepackage{psfrag}
\usepackage{epsfig}
\usepackage{epsf}
\usepackage{float}
\usepackage{slashed}

\newcommand{\beq}{\begin{equation}}
\newcommand{\eeq}{\end{equation}}
\newcommand{\beqa}{\begin{eqnarray}}
\newcommand{\eeqa}{\end{eqnarray}}

\begin{document}

\title{Few-nucleon systems with state-of-the-art chiral nucleon-nucleon forces}

%\author{V. Bernard}
%\affiliation{Institut de Physique Nucl\'eaire, CNRS/Univ. Paris-Sud
%  11, (UMR 8608), F-91406 Orsay Cedex, France}

\author{S.~Binder}
\affiliation{Physics Division, Oak Ridge National Laboratory, Oak
  Ridge, TN 37831, USA}

\author{A.~Calci}
\affiliation{TRIUMF, 4004 Wesbrook Mall, Vancouver, British Columbia, V6T 2A3 Canada}

\author{E.~Epelbaum}
\affiliation{Institut f\"ur Theoretische Physik II, Ruhr-Universit\"at
  Bochum, D-44780 Bochum, Germany}

\author{R.J.~Furnstahl}
\affiliation{Department of Physics, The Ohio State University, 
Columbus, Ohio 43210, USA}

\author{J.~Golak}
\affiliation{M. Smoluchowski Institute of Physics, Jagiellonian
University,  PL-30348 Krak\'ow, Poland}

\author{K.~Hebeler}
\affiliation{Institut f\"ur Kernphysik, Technische Universit\"at 
Darmstadt, 64289 Darmstadt, Germany}

\author{H.~Kamada}
\affiliation{Department of Physics, Faculty of Engineering,
Kyushu Institute of Technology, Kitakyushu 804-8550, Japan}

\author{H.~Krebs}
\affiliation{Institut f\"ur Theoretische Physik II, Ruhr-Universit\"at
  Bochum, D-44780 Bochum, Germany}

\author{J.~Langhammer}
\affiliation{Institut f\"ur Kernphysik, Technische Universit\"at 
Darmstadt, 64289 Darmstadt, Germany}

\author{S.~Liebig}
\affiliation{Institut f\"ur Kernphysik, Institute for Advanced Simulation 
and J\"ulich Center for Hadron Physics, Forschungszentrum J\"ulich, 
D-52425 J\"ulich, Germany}

\author{P.~Maris}
\affiliation{Department of Physics and Astronomy, Iowa State
  University, Ames, Iowa 50011, USA}

\author{Ulf-G.~Mei{\ss}ner}
\affiliation{Helmholtz-Institut~f\"{u}r~Strahlen-~und~Kernphysik~and~Bethe~Center~for~Theoretical~Physics,
~Universit\"{a}t~Bonn,~D-53115~Bonn,~Germany}
\affiliation{Institut f\"ur Kernphysik, Institute for Advanced Simulation 
and J\"ulich Center for Hadron Physics, Forschungszentrum J\"ulich, 
D-52425 J\"ulich, Germany}
\affiliation{JARA~-~High~Performance~Computing,~Forschungszentrum~J\"{u}lich,~D-52425~J\"{u}lich,~Germany}

\author{D.~Minossi}
\affiliation{Institut f\"ur Kernphysik, Institute for Advanced Simulation 
and J\"ulich Center for Hadron Physics, Forschungszentrum J\"ulich, 
D-52425 J\"ulich, Germany}

\author{A.~Nogga}
\affiliation{Institut f\"ur Kernphysik, Institute for Advanced Simulation 
and J\"ulich Center for Hadron Physics, Forschungszentrum J\"ulich, 
D-52425 J\"ulich, Germany}

\author{H.~Potter}
\affiliation{Department of Physics and Astronomy, Iowa State
  University, Ames, Iowa 50011, USA}

\author{R.~Roth}
\affiliation{Institut f\"ur Kernphysik, Technische Universit\"at 
Darmstadt, 64289 Darmstadt, Germany}

\author{R.~Skibi\'nski}
\affiliation{M. Smoluchowski Institute of Physics, Jagiellonian
University,  PL-30348 Krak\'ow, Poland}

\author{K.~Topolnicki}
\affiliation{M. Smoluchowski Institute of Physics, Jagiellonian
University,  PL-30348 Krak\'ow, Poland}

\author{J.P.~Vary}
\affiliation{Department of Physics and Astronomy, Iowa State
  University, Ames, Iowa 50011, USA}

\author{H.~Wita{\l}a}
\affiliation{M. Smoluchowski Institute of Physics, Jagiellonian
University,  PL-30348 Krak\'ow, Poland}

\date{\today}

\begin{abstract}
We apply improved nucleon-nucleon potentials up to fifth order in chiral
        effective field theory, along with a new analysis of the theoretical
        truncation errors, to study nucleon-deuteron (Nd) scattering and selected
        low-energy observables in $^3$H, $^4$He, and $^6$Li. Calculations beyond
        second order differ from experiment well outside the range of quantified
        uncertainties, providing truly unambiguous evidence for missing
        three-nucleon forces within the employed framework. The sizes of the
        required three-nucleon force contributions agree well with expectations
        based on Weinberg's power counting. We identify the energy range in
        elastic Nd scattering best suited to study three-nucleon force effects
        and estimate the achievable accuracy of theoretical predictions for
        various observables.
\end{abstract}

\maketitle
Chiral effective field theory (EFT) provides a powerful framework for
analyzing low-energy nuclear structure and reactions in harmony with the symmetries of
quantum chromodynamics (QCD), the underlying theory of the strong
interactions. It allows one to derive nuclear forces and currents in a systematically
improvable way in terms of a perturbative expansion in powers of
$Q \in  (p/\Lambda_b, M_\pi/\Lambda_b )$, the so-called chiral
expansion. Here, $p$ refers to the magnitude of 
the nucleon three-momentum, 
$M_\pi$ is the pion mass and $\Lambda_b$ is the breakdown
scale of chiral EFT \cite{Weinberg:1990rz}. One finds, in particular,
that the leading-order (LO) contribution to the Hamiltonian at
order $Q^0$ and the first corrections at order $Q^2$ (NLO) are given
solely by nucleon-nucleon (NN) operators while three-nucleon forces (3NFs) appear first at
order $Q^3$  (N$^2$LO) (see \cite{Epelbaum:2008ga} and references therein). 
Four-nucleon forces are even
more suppressed and start contributing at order $Q^4$
(N$^3$LO). The chiral power counting thus provides a
natural explanation of the observed hierarchy of nuclear forces. 

With accurate N$^3$LO NN potentials being available since about a
decade \cite{Entem:2003ft,Epelbaum:2004fk}, the main focus of research
has moved in recent years towards 
the 3NF \cite{KalantarNayestanaki:2011wz,Hammer:2012id}.  While providing a small corrections to the nuclear
Hamiltonian as compared to the dominant NN force, its inclusion is
mandatory for quantitative understanding of nuclear structure and 
reactions. Historically, the importance of the 3NF has been pointed out
already in the thirties \cite{Primakoff:1939zz} while the first phenomenological 3NF models
date back to the fifties. However, in spite of
extensive efforts, the spin structure of the 3NF is still poorly
understood \cite{KalantarNayestanaki:2011wz}.

Chiral EFT is expected to provide a suitable theoretical resolution to the
long-standing 3NF problem. Indeed, the leading
chiral 3NF has already been extensively explored in \emph{ab initio} calculations by various
groups and found to yield promising results for nuclear structure
and reactions \cite{Hammer:2012id,Barrett:2013nh}.  
The first corrections to the 3NF at order $Q^4$ (N$^3$LO)
have also been derived \cite{Ishikawa:2007zz,Bernard:2007sp,Bernard:2011zr} (and appear to be parameter-free) while
the sub-subleading contributions at order $Q^5$ (N$^4$LO)
are being derived \cite{Girlanda:2011fh,Krebs:2012yv,Krebs:2013kha}. 

On the other hand, understanding and validating the fine details of the 3NF 
%which itself is a small correction to the dominant NN interaction,
clearly requires precise and systematic NN potentials and 
a reliable approach for estimating the accuracy of theoretical predictions at a
given chiral order. 
Refs.~\cite{Epelbaum:2014efa,Epelbaum:2014sza}
initiated the direction that we follow here by developing a new
generation of chiral EFT NN forces up to N$^4$LO, in which the amount of
finite-cutoff artefacts has been substantially reduced by employing a
novel ultraviolet regularization scheme, and by introducing a new  
procedure for estimating the theoretical uncertainty.

In this Letter we,
for the first time, apply these novel chiral NN forces 
beyond the two-nucleon system and demonstrate their suitability  for
modern \emph{ab initio} few- and many-body methods.  By applying the   
new method for error analysis, we present unambiguous evidence for
missing 3NF effects 
and demonstrate that the size of
the required 3NF contributions agrees well with expectations
based on Weinberg's power counting. We also estimate
the theoretical accuracy for various observables achievable at N$^4$LO and identify
the energy region in elastic Nd scattering that is best suited for
testing the chiral 3NF. 

We first describe our procedure for estimating the theoretical
uncertainty. 
Let $X(p)$ be some observable with $p$ referring to the corresponding momentum scale and $X^{(i)} (p)$, $i = 0, 2, 3, \ldots$, 
a prediction at order $Q^i$ in the chiral expansion. 
We 
further define the order-$Q^i$ corrections to $X(p)$ via 
\begin{eqnarray}
\Delta X^{(2)} &\equiv& X^{(2)} - X^{(0)}, \nonumber \\ 
\Delta X^{(i)} &\equiv& X^{(i)} - X^{(i-1)}, \;\; i \geq 3 \,,
\end{eqnarray}  
so that the chiral expansion for $X$ takes the form  
\begin{equation}
X^{(i)} = X^{(0)}+ \Delta X^{(2)}+ \ldots +  \Delta X^{(i)}\,.
\end{equation}  
Generally, the size of the corrections is expected to be 
\begin{equation}
\label{size}
\Delta X^{(i)} = \mathcal{O} ( Q^i X^{(0)}). 
\end{equation}
In \cite{Epelbaum:2014sza}, the validity of this estimation was
confirmed for the total neutron-proton cross section.  
In Refs. ~\cite{Epelbaum:2014efa,Epelbaum:2014sza}, quantitative estimates of the theoretical uncertainty
      $\delta X^{(i)}$ of the chiral EFT prediction $X^{(i)}$ were made using the
      expected and actual sizes of higher-order contributions.
%In Refs.~\cite{Epelbaum:2014efa,Epelbaum:2014sza}, the theoretical uncertainty $\delta X^{(i)}$ of the chiral EFT prediction $X^{(i)}$ was estimated in terms of  
%the expected size of higher-order contributions. 
Specifically, the following procedure was employed:
\begin{eqnarray}
\label{Err}
\delta X^{(0)} &=& Q^2 | X^{(0)} |,  \\
\delta X^{(i)} &=& \max \Big( Q^{i+1} | X^{(0)} |, \,  Q^{i+1-j} | \Delta X^{(j)} | \Big),  \;  2 \leq j \leq i   
\nonumber
\end{eqnarray}
where $i \geq 2$ and $Q = \max  (p/\Lambda_b, M_\pi/\Lambda_b )$ with 
$\Lambda_b = 600$, $500$ and $400\;$MeV for the regulator
choices of $R=0.8 - 1.0\;$fm, $R=1.1\;$fm and $R=1.2\;$fm,
respectively.
The sizes of actual higher-order calculations provide additional
       information on the theoretical uncertainties, which we use by adding
       the conditions
\begin{equation}
\label{ErrExplicit}
\delta X^{(i)} \,  \ge \,   \max \Big( \big| X^{(j \ge i)} -  X^{(k \ge i)} \big| \Big)
\end{equation}
 to estimates of lower-order uncertainties.

The above procedure for estimating the uncertainty needs to be adjusted in order 
to account for the neglect of many-body forces in the present analysis. 
In particular, iterating the NN T-matrix in the Faddeev equation
generates contributions whose short-range behavior is order- and regulator-dependent.
%*******************************
For low-energy Nd observables calculated in the EFT framework, approximate scheme
independence is restored upon performing
renormalization, i.e.~upon 
expressing the bare low-energy constants (LECs) accompanying short-range 3NFs at orders $Q^3, \,Q^5, \, \ldots$ in terms 
of observable quantities, such as the triton binding
energy. In practice, this is achieved by fitting the
corresponding LECs to experimental data. 
Therefore, when performing incomplete calculations 
based on NN interactions only, the estimation  in Eq.~(\ref{size}) is not
justified at or beyond N$^2$LO, the chiral order at which the 
contact 3NF starts contributing. We, therefore, adopt here a slightly
modified procedure for estimating the uncertainty $\delta X^{(i)}$
for $i \ge 3$, namely 
\begin{equation}
\label{ErrMod}
\delta X^{(i)} = \max \Big( Q^{i+1} | X^{(0)} |, \,  Q^{i-1} |
\Delta X^{(2)} |, \,  Q^{i-2} | \Delta X^{(3)} | \Big),    
\nonumber
\end{equation}
and do not employ  Eq.~(\ref{ErrExplicit}). However, to be conservative in our estimates, we further require that
\begin{equation}
\label{ErrExplicitMod}
\delta X^{(2)}  \ge Q \, \delta X^{(0)}, \quad \delta X^{(i \ge 3)}
\ge Q \, \delta X^{(i-1)}\,.
\end{equation}

The dependence of the chiral NN forces on the local regulator $R$ over the range 
$0.8 \ldots 1.2$ fm has been extensively investigated in Ref.~\cite{Epelbaum:2014sza} 
showing that cutoff artifacts become visible for $R > 1.0$ fm.  On the other hand, we seek to obtain many-body results as close
to convergence as possible, and this favors the largest feasible value of $R$.  We therefore balance these 
competing conditions with the choice of $R = 1.0$ fm in this work.

\begin{figure}[tb]
\includegraphics[width=0.49\textwidth,keepaspectratio,angle=0,clip]{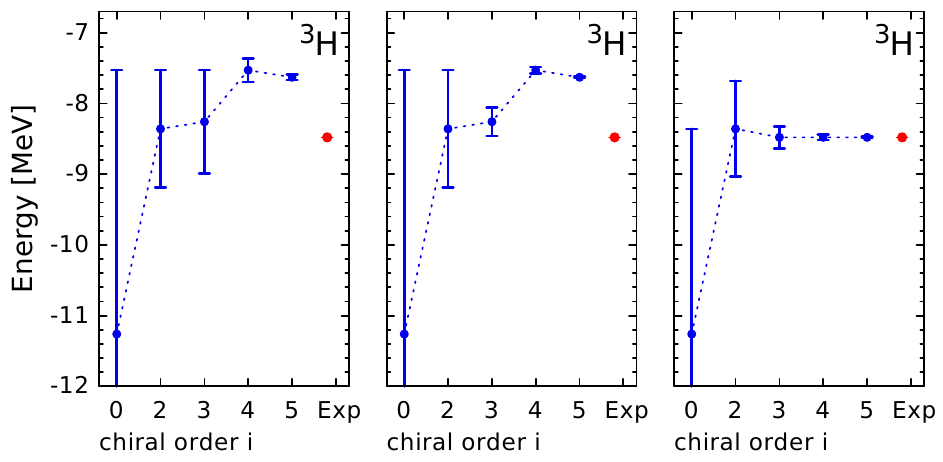}
\caption{(Color online) Chiral expansion of the $^3$H $E_{\rm gs}$ based on the
  NN potentials of Refs.~\cite{Epelbaum:2014efa,Epelbaum:2014sza} for the regulator $R=1.0$
  fm and using $Q=M_\pi/\Lambda_b$. Left (middle) panel shows incomplete results based on NN
  forces only, with uncertainties being estimated via
  Eqs.~(\ref{Err}, \ref{ErrExplicit}) (Eqs.~(\ref{ErrMod}, \ref{ErrExplicitMod}) for $i \ge 3$). Right panel
  shows the projected results
 % complete result 
  assuming that the LECs in the N$^2$LO 3NF
  are tuned to reproduce the $^3$H $E_{\rm gs}$ and using
  Eqs.~(\ref{Err}, \ref{ErrExplicit}) to specify the uncertainty. 
\label{Fig:3H} 
 }
\end{figure}

Our results for the chiral expansion of the $^3$H 
%binding energy (BE)
ground state energy ($E_{\rm gs}$)
using $Q=M_\pi / \Lambda_b$ are visualized in
Fig.~\ref{Fig:3H}. 
Assuming that the LECs which enter the short-range part of the 
%N$^2$LO
3NF can be tuned to reproduce the $^3$H $E_{\rm gs}$, we can already at this
stage present a \emph{complete} result up to N$^4$LO for the chiral expansion of
this observable, see the right panel in  Fig.~\ref{Fig:3H}. As
expected, we observe that Eqs.~(\ref{ErrMod}, \ref{ErrExplicitMod})
provide a more reliable approach for error estimation in
calculations based on NN interactions only, while using
Eqs.~(\ref{Err}, \ref{ErrExplicit}) amounts to overestimating 
the actual error.
The 
N$^3$LO (N$^4$LO) results for the $^3$H $E_{\rm gs}$ are expected to be accurate
at the level of ${\sim}50\;$keV (${\sim}10\;$keV) for the regulator 
choices of $R=0.8$, $0.9$ and $1.0\;$fm.
%It is reassuring to see
Note
that the size of the 3NF contribution agrees well with the uncertainty at NLO, 
which reflects the estimated impact of the N$^2$LO contributions to the
Hamiltonian. This is fully in line with expectations based on the
Weinberg power counting \cite{Weinberg:1990rz,Epelbaum:2008ga}. 

\begin{figure}[tb]
\includegraphics[width=0.49\textwidth,keepaspectratio,angle=0,clip]{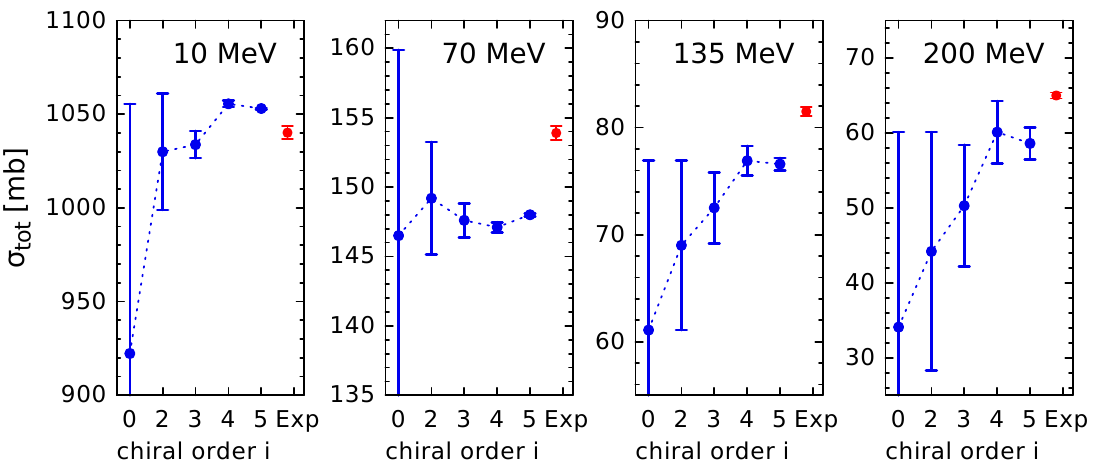}
    \caption{(Color online) Predictions for Nd total cross section based on the
  NN potentials of Refs.~\cite{Epelbaum:2014efa,Epelbaum:2014sza} for $R=1.0\;$fm
  without including the 3NF. Theoretical uncertainties (blue) 
  are estimated via Eqs.~(\ref{Err}) and (\ref{ErrExplicit}) 
  for chiral order $i = 0, 2$ and via Eqs.~(\ref{ErrMod}) and (\ref{ErrExplicitMod}) for $i \ge 3$. 
  Experimental data are
  from Ref.~\cite{Abfalterer:1998zz}. 
\label{Fig:NdTot} 
 }
\end{figure}

We now turn to Nd scattering observables. Our predictions for the
Nd total cross section are visualized in Fig.~\ref{Fig:NdTot}.   
Similar to the $^3$H $E_{\rm gs}$, one observes a significant discrepancy between
the theoretical predictions based on the NN forces only and data, 
which provides clear evidence for
missing 3NF contributions. 
%Except for the lowest energy, 
The size of
the discrepancy agrees within 1.5 times
the estimated size of N$^2$LO
corrections shown by the NLO error bars. Interestingly,  the
discrepancy at the lowest energy of $10\;$MeV is much smaller than the
estimated size of N$^2$LO contributions. Given that the cross section at
low energy is governed by the S-wave spin-doublet and spin-quartet Nd
scattering lengths, this observation can be naturally explained. Indeed,
the spin-quartet scattering length is almost an order of magnitude
larger than that of the spin-doublet and much less sensitive to the 3NF 
as a consequence of the Pauli principle.  

\begin{figure}[tb]
\includegraphics[width=0.49\textwidth,keepaspectratio,angle=0,clip]{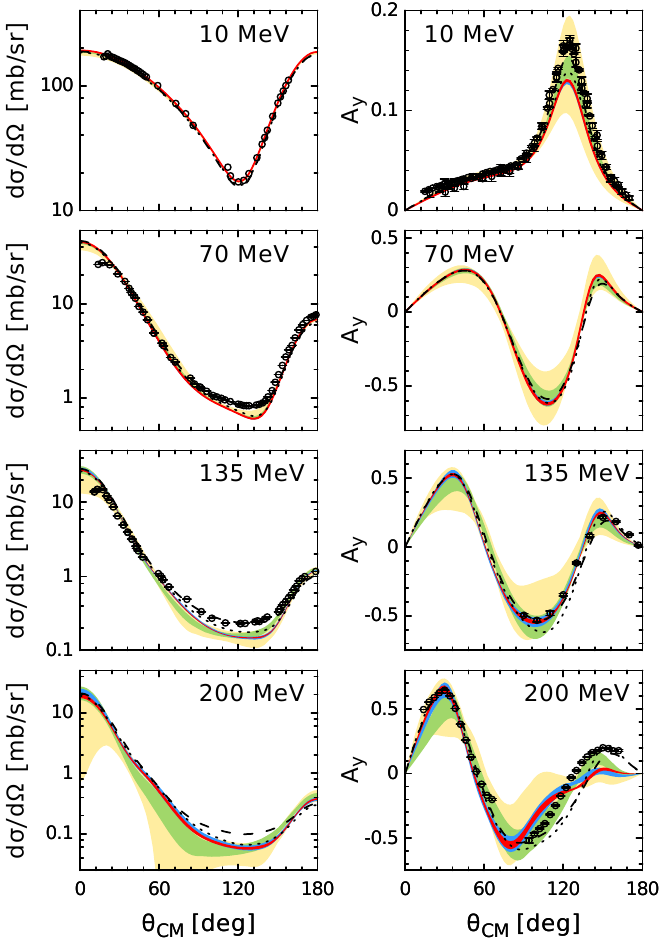}
    \caption{(Color online) Predictions for the differential cross section and nucleon
    $A_y$ in elastic Nd scattering based on the
  NN potentials of Refs.~\cite{Epelbaum:2014efa,Epelbaum:2014sza} for $R=1.0\;$fm
  without including the 3NF. Theoretical uncertainties  
  are estimated via Eqs.~(\ref{Err}) and (\ref{ErrExplicit}) 
  for chiral order $i = 2$ and via Eqs.~(\ref{ErrMod}) and (\ref{ErrExplicitMod}) for $i \ge 3$. 
  The bands of increasing
  width show estimated theoretical uncertainty at N$^4$LO (red),
  N$^3$LO (blue), N$^2$LO (green) and NLO
  (yellow). The dotted (dashed) lines show the results
  based on the CD Bonn NN potential \cite{Machleidt:2000ge} (CD Bonn NN potential in
  combination with the Tucson-Melbourne 3NF \cite{Coon:2001pv}). For
  references to proton-nucleon data (symbols) see Ref.~\cite{KalantarNayestanaki:2011wz}. 
\label{Fig:Nd1} 
 }
\end{figure}

\begin{figure}[tb]
\includegraphics[width=0.49\textwidth,keepaspectratio,angle=0,clip]{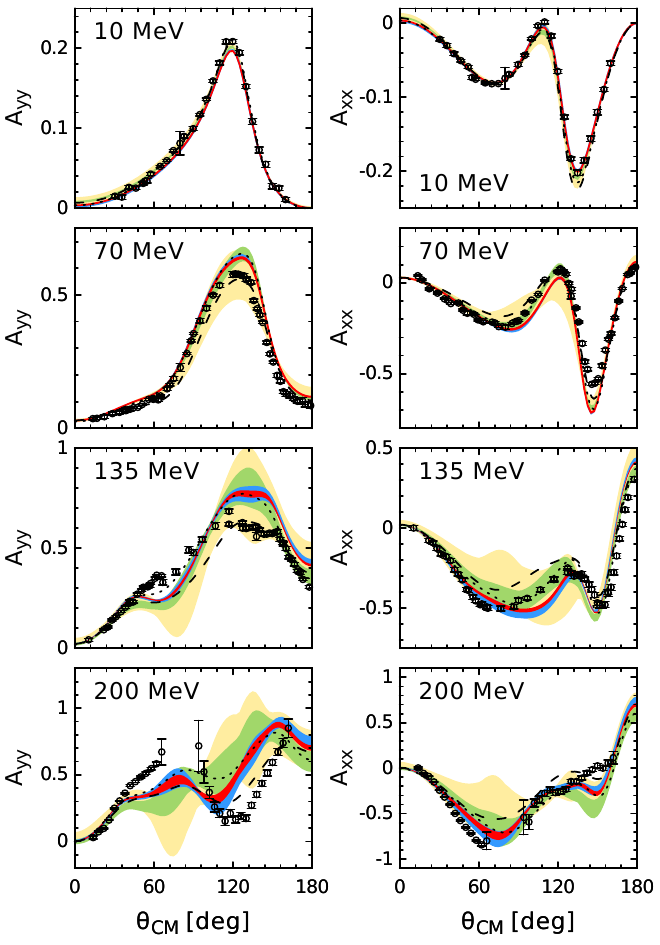}
    \caption{(Color online) Predictions for the tensor analyzing powers $A_{yy}$
      and $A_{xx}$  in elastic Nd scattering based on the
  NN potentials of Refs.~\cite{Epelbaum:2014efa,Epelbaum:2014sza} for $R=1.0$
  fm without including the 3NF. For notations see Fig.~\ref{Fig:Nd1}. 
\label{Fig:Nd2} 
 }
\end{figure}

Our predictions for Nd differential cross section and analyzing
powers $A_y (N)$, $A_{yy}$ and $A_{xx}$ are shown in
Figs.~\ref{Fig:Nd1},  \ref{Fig:Nd2}.
At the lowest energy of $10\;$MeV, there is little apparent need for 3NF
effects except for $A_y$. Interestingly, the fine-tuning nature of
this observable is clearly reflected in large theoretical
uncertainties at NLO and N$^2$LO. Starting from $E_{N} = 70\;$MeV, one
observes clear discrepancies between our predictions and data for the
cross section and tensor analyzing powers which are expected to be
explained by the 3NF. In all cases, the required 3NF
contributions are of a natural size. Based on the width of the bands,
one may expect Nd scattering observables at N$^4$LO
to be accurately described up to energies of 
at least $200\;$MeV. It is also comforting to see
that the accuracy of chiral EFT predictions for Nd and NN \cite{Epelbaum:2014sza} scattering
observables at the same energy is comparable. We further emphasize
that the improved NN potentials of Refs.~\cite{Epelbaum:2014efa,Epelbaum:2014sza} show clearly 
smaller finite-cutoff artifacts as compared to the N$^3$LO potentials
of Refs.~\cite{Entem:2003ft,Epelbaum:2004fk} and, in particular, do
not lead to distortions in the cross section minimum that were found in
Ref.~\cite{Witala:2013ioa}.

\begin{figure}[tb]
\includegraphics[width=0.99\columnwidth,keepaspectratio,angle=0,clip]{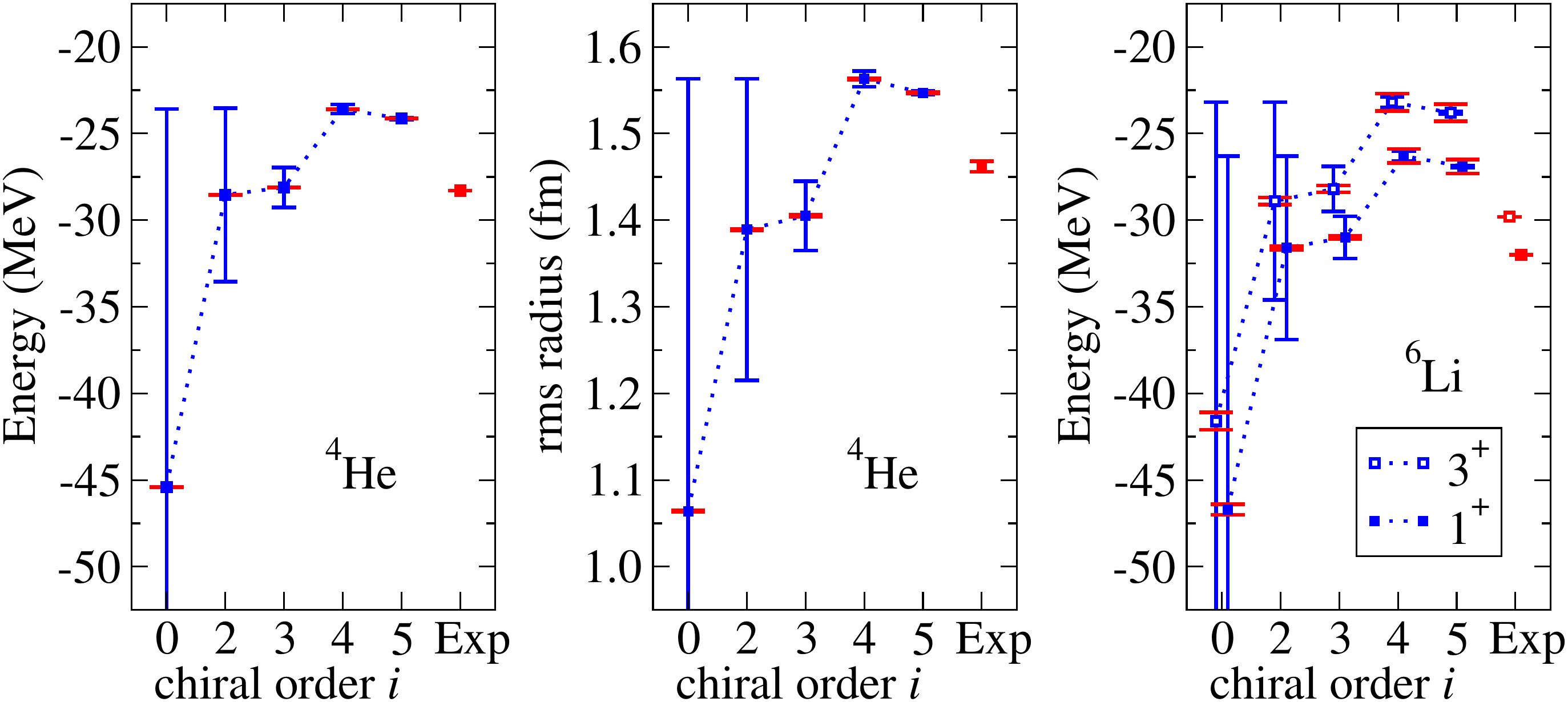}
    \caption{(Color online) Predictions for $E_{\rm gs}$ and $r_{\rm p}$ of $^4$He 
    and the energies of the lowest two states of $^6$Li based on the
  NN potentials of Refs.~\cite{Epelbaum:2014efa,Epelbaum:2014sza} for $R=1.0$
  fm without including the 3NF. Theoretical uncertainties (blue) 
  are estimated via Eqs.~(\ref{Err}) and (\ref{ErrExplicit}) 
  for chiral order $i = 0, 2$ and via Eqs.~(\ref{ErrMod}) and (\ref{ErrExplicitMod}) for $i \ge 3$.
 Numerical uncertainties from the NCSM
%many-body methods 
 (red) are estimated following 
%  Refs. [Andreas?] and 
Ref.~\cite{Maris:2013poa}.
\label{Fig:A4_6}
 }
\end{figure}

Next, we apply the improved NN potentials to $A > 3$ systems.
We present in Fig.~\ref{Fig:A4_6} order-by-order calculations of
selected observables for $^4$He and $^6$Li.  The results for $^4$He are
obtained both by solving the Faddeev-Yakubovsky (FY) equations and
with the no-core shell model (NCSM) \cite{Barrett:2013nh}, which agree to within the
estimated uncertainties 
of these methods.  The numerical uncertainties in the FY solutions
are a few keV for the energy and about $0.001$ fm for the point-proton
radius ($r_p$).  The numerical uncertainties from incomplete
convergence of the NCSM (see Ref.~\cite{Maris:2013poa} for details) are shown as error
bars (color online: red) together with the estimated theoretical uncertainties from
the truncated chiral expansion with $Q = M_\pi/\Lambda_b$ (color
online: blue). 

%Next, we apply the improved NN potentials to $A>3$ systems, 
%where we first confront  
%practical considerations for the convergence of solution
%methods such as the no-core shell model (NCSM)~\cite{Barrett:2013nh}.
%In particular, the question arises if the new regularization scheme might lead
%to impractically hard interactions. A quantitative
%metric of the softness of NN potentials is given by Weinberg 
%eigenvalues~\cite{Weinberg:1963zz,Bogner:2006tw}. 
%While a detailed analysis will be presented elsewhere, we merely note
%that Weinberg eigenvalues for the new chiral potentials
%are comparable in magnitude to those from previous N$^3$LO
%interactions with similar cutoffs~\cite{Epelbaum:2004fk}.

%We present in Fig. \ref{Fig:A4_6} selected observables for $^4$He and $^6$Li 
%with the numerical
%uncertainty arising from 
%incomplete convergence of the many-body methods, see
%Ref.~\cite{Maris:2013poa} for details.  
%The results for $^4$He
%are obtained both with the NCSM and by solving the Faddeev-Yakubovsky
%(FY) equations 
%%[Andreas?] 
%which agree to within their numerical uncertainties.
%The numerical uncertainties in these FY solutions are about a few 
%%1(Andreas?) 
%keV for the energy and 0.001(Andreas?) fm 
%for the point-proton rms radius ($r_{\rm p}$). 

For the $^6$Li
energies, we carried 
out Similarity Renormalization Group (SRG) evolution \cite{Bogner:2007rx} in order 
to enhance the convergence rate of the NCSM calculations that were
performed in basis  spaces up through $N_{\rm max}=12$ and
extrapolated to the infinite matrix limit 
following Ref.~\cite{Bogner:2007rx}. We retained the induced 3NF arising from the SRG 
evolution, see Ref.~\cite{Roth:2010bm} for details, and this produces results for the $^6$Li energies in Fig.~\ref{Fig:A4_6}
that are independent of the SRG scale over the range $\alpha = 0.04 - 0.08$~fm$^4$ 
to within our quoted many-body uncertainties. For example, at N$^4$LO we obtain 
$E_{\rm gs} = -26.9(4)$ $(-26.9(2))$~MeV at $\alpha = 0.04$($0.08$)~fm$^4$
for $^6$Li where the quantified numerical uncertainty in the last digit 
of the energy is quoted in parenthesis.

The patterns for the 
energies in Fig.~\ref{Fig:A4_6} as well as for the $r_{\rm p}$ of $^4$He are
very similar to the pattern for the $E_{\rm gs}$ of $^3$H in
Fig.~\ref{Fig:3H} and the Nd total cross section at $10$ MeV in
Fig.~\ref{Fig:NdTot}.  As in $^3$H, we 
again observe underbinding indicative of the need for 3NFs, especially
at N$^3$LO and N$^4$LO.  
This underbinding is correlated with larger $r_{\rm p}$ in
 $^4$He, which is expected to decrease toward the experimental result as
 $E_{\rm gs}$ is lowered toward experiment with the inclusion
 of 3NFs. Note that the energy of the first excited state in $^6$Li, with 
$J^\pi=3^+$, follows the same pattern as the ground state energy, 
leading to an excitation energy that depends much less on the chiral 
order than one might naively expect based on the theoretical 
uncertainties of the binding energies.

To summarize, we have studied in this Letter  selected few-nucleon observables using
improved chiral NN potentials of
Refs.~\cite{Epelbaum:2014efa,Epelbaum:2014sza} up to N$^4$LO. 
Our results suggest that these new chiral forces are well 
%We have demonstrated that these new chiral forces are well
suited for modern \emph{ab initio} few- and many-body methods. Using
the novel approach for error analysis introduced in Ref.~\cite{Epelbaum:2014efa}, we
found truly unambiguous evidence for missing 3NF effects by observing
discrepancies between our predictions and experimental data well outside 
the range of quantified uncertainties. The magnitude of these
discrepancies is found to match well with the
expected size of the chiral 3NF whose dominant contribution appears at
N$^2$LO. Furthermore, we have
demonstrated that the predictions for Nd and NN scattering observables at the same energy
have comparable accuracy, in agreement with the general principles of
EFT. Most importantly, the expected theoretical uncertainty for Nd
scattering observables at N$^3$LO and N$^4$LO in the energy range
of $E_{\rm   lab } \simeq 70 - 200$ MeV is shown to be substantially smaller than
the observed discrepancies between state-of-the-art calculations and
experimental data.  This suggests
that chiral EFT at these orders should be capable of resolving the long
standing 3NF problem in nuclear physics. Work on the explicit
inclusion of the consistent 3NFs is in progress.

This work was performed by the LENPIC collaboration with support from: 
BMBF (contract 06DA7047I);
DFG and NSFC (CRC 110); DFG (SFB 634, SFB/TR 16);
ERC projects 259218 NUCLEAREFT and 307986 STRONGINT;
EU (HadronPhysics3, Grant 283286);
HIC for FAIR;
Polish National Science Center DEC-2013/10/M/ST2/00420;
US DOE DESC0008485, DE-FG-02-87ER40371, DE-SC0006758, DE-SC0008533;
US NSF PHY-0904782, PHY-1306250.
Supercomputer usage included:
JSC-J\"ulich; 
PAS0680-Ohio SC;
Edison-NERSC.

\end{document}